\documentclass[pra,aps,twocolumn,superscriptaddress]{revtex4}
\usepackage{hyperref}
\usepackage{bm}
\usepackage{epsfig}
\usepackage{amssymb}
\usepackage{graphicx}

\usepackage{amsmath}
\usepackage{epsfig}

\begin{document}
\title{Non-unity gain minimal disturbance measurement}
\author{Metin Sabuncu}
\affiliation{Department of Physics,
Technical University of Denmark, 2800 Kongens Lyngby, Denmark}
\affiliation{Institut f\"{u}r Optik, Information und Photonik,
Max-Planck Forschungsgruppe, Universit\"{a}t
Erlangen-N\"{u}rnberg, G\"{u}nther-Scharowsky str. 1, 91058,
Erlangen, Germany} 
\author{Ladislav Mi\v{s}ta, Jr.}
\affiliation{Department of Optics, Palack\' y University, 17.
listopadu 50,  772~07 Olomouc, Czech Republic} \affiliation{School
of Physics and Astronomy, University of St. Andrews, North Haugh,
St. Andrews, Fife, KY16 9SS, Scotland}
\author{Jarom\'{\i}r Fiur\'a\v{s}ek}
\affiliation{Department of Optics, Palack\' y University, 17.
listopadu 50,  772~07 Olomouc, Czech Republic}
\author{Radim Filip}
\affiliation{Institut f\"{u}r Optik, Information und Photonik,
Max-Planck Forschungsgruppe, Universit\"{a}t
Erlangen-N\"{u}rnberg, G\"{u}nther-Scharowsky str. 1, 91058,
Erlangen, Germany}
\affiliation{Department of Optics, Palack\' y
University, 17. listopadu 50,  772~07 Olomouc, Czech Republic}
\author{Gerd Leuchs}
\affiliation{Institut f\"{u}r Optik, Information und Photonik,
Max-Planck Forschungsgruppe, Universit\"{a}t
Erlangen-N\"{u}rnberg, G\"{u}nther-Scharowsky str. 1, 91058,
Erlangen, Germany}
\author{Ulrik L. Andersen}
\affiliation{Department of Physics,
Technical University of Denmark, 2800 Kongens Lyngby, Denmark}
\affiliation{Institut f\"{u}r Optik,
Information und Photonik, Max-Planck Forschungsgruppe,
Universit\"{a}t Erlangen-N\"{u}rnberg, G\"{u}nther-Scharowsky str.
1, 91058, Erlangen, Germany} 
\date{\today}


\date{\today}

\begin{abstract}
We propose and experimentally demonstrate an optimal non-unity gain
Gaussian scheme for partial measurement of an unknown coherent
state that causes minimal disturbance of the state. The
information gain and the state disturbance are quantified by the
noise added to the measurement outcomes and to the output state,
respectively. We derive the optimal trade-off relation between the two noises
and we show that the trade-off is saturated by  non-unity gain
teleportation. Optimal partial measurement is demonstrated
experimentally using a linear optics scheme with  feed-forward.
\end{abstract}

\maketitle
\section{Introduction}

One of the most counter-intuitive concepts of quantum mechanics is
the fact that any attempt to gain information on an unknown
quantum state of a physical system will inevitably result in a
noisy feedback to the measured system. No matter how cleverly the
measurement is performed, the state will always be disturbed to
some extent: The more information obtained about a quantum
state from a measurement, the more it will be altered, and vice
versa. Although this measurement-disturbance concept is very old
and originally only of fundamental interest, it has recently
received renewed interest due to its direct application in the
flourishing field of quantum information science, and in
particular, quantum key distribution.

The study of the interplay between the quality of the estimation of a
quantum state and the disturbance of the post-measurement state has been
extensively carried out in finite-dimensional systems, where
optimal trade-off relations have been established for various
cases~\cite{Banaszek_01a, Banaszek_01b,Mista_05,Mista_05b,sacchi06,barnum02,fuchs96,maccone06} and realized
recently in an experiment ~\cite{Sciarrino_05}.
In contrast, much less effort has been devoted to
the study of this trade-off in infinitely-dimensional systems~\cite{Andersen_06,mista06.pra,genoni,olivares}
where quantum information is carried by observables with a
continuous spectrum, important examples being the canonically
conjugate quadrature amplitudes. Gaussian states which belong to
continuous variable states have played a key role in various
experimental realizations of quantum information protocols, thanks
to the ease in generating and handling them in a quantum optics
lab~\cite{book,braunstein05.rev}. In the Gaussian scenario, full
control of the trade-off between the quality of measurement and
state disturbance was recently demonstrated for coherent states using a simple scheme
relying solely on linear optics and homodyne detection and near optimal performance
was reported~\cite{Andersen_06}. 

\begin{figure}
\centerline{\psfig{width=8.0cm,angle=0,file=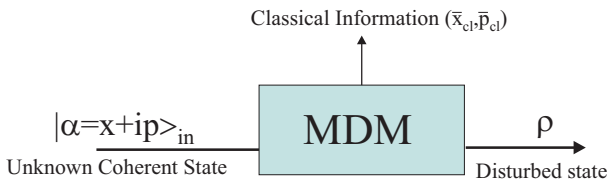}}
\caption{The principles of a minimal disturbance measurement of coherent states. The input state is drawn from an unknown distribution of coherent states, say $|\alpha\rangle_{in} =|x+ip\rangle_{in}$, and the task is to acquire information about the state (through a measurement) in such a way that the state is minimally disturbed according to quantum mechanics. This is the essence of a minimal disturbance measurement. There are two outputs of the protocol; a classical one yielding information about the intput state (in form of two numbers, say $\bar x_{cl}$ and $\bar p_{cl}$) and a quantum one, namely the post measurement state $\rho$. In this paper we consider only cases where the classical data as well as the disturbed quantum state are inflicted by additive phase insensitive Gaussian noise. The gain, $g$, of the protocol is defined by the ratio between the input and output mean values: $g=tr(x\rho)/\langle\alpha_{in}|x|\alpha_{in}\rangle=tr(p\rho)/\langle\alpha_{in}|p|\alpha_{in}\rangle$.}
\label{fig0}
\end{figure}

Let us define the problem that will be addressed in this
paper. The task is to perform a minimal disturbance measurement on
a coherent state which is taken from an unknown distribution (see Fig.~\ref{fig0}). That is, a completely random
coherent state will be received by our measurement device. The
question that will be raised and answered in this paper is: What
is the optimal information disturbance trade-off for this scenario? The answer to that
question depends on the figure of merit used to quantify the
information gain and the measurement disturbance. For Gaussian states, a useful and practical measure of the quality of the measurement is the phase insensitive added noise~\cite{Poizat_94}, since it directly determines the Shannon information optimally extracted by the measurement. Thus the
optimal trade-off between the added noises determines 
the maximal information that can be gained from the Gaussian measurement represented
by a channel with a given additive noise. A second parameter of high relevance for describing the measurement is the gain (attenuation or amplification) of the channel, since the minimization of the added noise is done with respect to that gain. For example in the previous experiment on minimal disturbance measurement~\cite{Andersen_06}, the added noise was minimized under the constraint that the channel gain was unity (corresponding to a conservation of the mean values). For Gaussian measurements and Gaussian channels this optimization procedure corresponds to a maximization of the fidelity over all possible input states drawn from the unknown coherent state alphabet. It should however be noted that by using the fidelity as a measure the optimal solution is non-Gaussian~\cite{mista06.pra} due to specific properties of fidelity.

In this paper we investigate theoretically and experimentally the
optimal trade-off relations in terms of added noises using two different strategies. In the first approach the channel gain is a free parameter that is optimised to minimize the trade-off between the added noises associated with the measurement and disturbance. This trade-off relation was derived by Ralph \cite{Ralph_00.pra} who also found that the relation could be experimentally demonstrated employing an ideal teleportation scheme 
with tunable entanglement. Here we propose and experimentally realise a different approach which is not relying on entanglement but solely on linear optics, Gaussian measurements and feed-forward similar to the one employed in Ref.~\cite{Andersen_06}. 

In the second approach that will be carefully addressed in this paper, the channel gain of the minimal disturbance measurement is fixed to a certain value associated with a particular realisation (the unity gain operation demonstarted in Ref.~\cite{Andersen_06} being a special case). For this case we derive a trade-off relation for arbitrary gains and prove its optimality using two different complementary proofs. As in the previous case, we also find here that a scheme similar to the one in Ref.~\cite{Andersen_06} can be used to implement the optimal trade-off for fixed but non-unity gain operation. This is demonstrated and near optimal performance is achieved. The experimental scheme is not only of fundamental interest but it can be also applied to perform optimal individual Gaussian attacks in a continuous-variable quantum key distribution scheme based on heterodyne detection~\cite{Weedbrook_04,lorenz}. 

The paper is organized as follows. Section~\ref{sec_1} deals in general with
trade-off between added noises and in section~\ref{sec1b} the trade-off is exemplified by the non-unity gain
teleportation scheme. In Section~\ref{sec_2} we give two different proofs
of optimality of the trade-off. Section~\ref{sec_3} is dedicated
to linear optical scheme saturating the trade-off. The
experimental demonstration of the scheme is given in
Section~\ref{sec_4}. In Section~\ref{sec_6} we discuss the possibility of using the minimal disturbance measurement as an eavesdropping attack, and finally we conclude the paper.

\section{Gaussian minimal disturbance measurements}\label{sec_1}
We consider Gaussian quantum operation that acts on a single mode of an
optical field ``in'' described by the canonically conjugate amplitude
and phase
quadratures $x_{\rm in}$ and $p_{\rm in}$ ($[x_{\rm in},p_{\rm
in}]=2i$). We assume that
the output mode of the operation is characterized by a pair of
quadratures
$x_{\rm out},p_{\rm out}$ ($[x_{\rm out},p_{\rm out}]=2i$) related to
the input quadratures by the formulas
\begin{eqnarray}\label{output}
x_{\rm out}&=&g(x_{\rm in}+n_{\rm{out,x}}),\quad  p_{\rm out}=g(p_{\rm
in}+n_{\rm{out,p}}),
\end{eqnarray}
where the quantity $g>0$ is the gain of the operation. The operators
$n_{\rm{out,x}}$ and $n_{\rm{out,p}}$ are standard operators of noises
added to the input state. The operation also outputs
a pair of mutually commuting variables $x_{\rm cl}$ and $p_{\rm cl}$ that
depend
linearly on the input quadratures $x_{\rm in}$ and $p_{\rm in}$ and
that
therefore can be used for simultaneous measurement of these
quadratures.
These variables can be expressed, after a suitable scaling
transformation, as
\begin{eqnarray}\label{classical}
x_{\rm cl}=x_{\rm in}+n_{\rm{cl,x}},\quad p_{\rm cl}=p_{\rm
in}+n_{\rm{cl,p}},
\end{eqnarray}
and satisfy the commutation rules
\begin{eqnarray}\label{clcommutators}
[x_{\rm cl},p_{\rm cl}]&=&[x_{\rm out},x_{\rm cl}]=[x_{\rm out},p_{\rm
cl}]=\nonumber\\
&=&[p_{\rm out},x_{\rm cl}]=[p_{\rm out},p_{\rm cl}]=0.
\end{eqnarray}
The operators $n_{\rm{cl,x}}$ and $n_{\rm{cl,p}}$ describe noises added
to the
outcomes of simultaneous measurement of input quadratures $x_{\rm in}$
and
$p_{\rm in}$ by homodyne detection of the variables $x_{\rm cl}$ and
$p_{\rm cl}$.
Naturally, the operators $n_{\rm{out,x}}$ and $n_{\rm{cl,x}}$
($n_{\rm{out,p}}$ and $n_{\rm{cl,p}}$)
are independent of the input quadrature $x_{\rm in}$ ($p_{\rm in}$) and
hence
\begin{eqnarray}\label{nxpcommutators1}
[n_{\rm out,x},p_{\rm in}]&=&[n_{\rm cl,x},p_{\rm in}]=[n_{\rm
out,p},x_{\rm in}]=\nonumber\\
&=&[n_{\rm cl,p},x_{\rm in}]=0.
\end{eqnarray}
In addition, the gains of the operation are assumed to be fixed for all
input states, i.e.
$\langle x_{\rm out}\rangle/\langle x_{\rm in}\rangle=\langle p_{\rm
out}\rangle/\langle p_{\rm in}\rangle=g$,
$\langle x_{\rm cl}\rangle/\langle x_{\rm in}\rangle=\langle p_{\rm
cl}\rangle/\langle p_{\rm in}\rangle=1$,
which implies that
\begin{eqnarray}\label{nxpcommutators2}
[n_{\rm out,x},x_{\rm in}]&=&[n_{\rm out,p},p_{\rm in}]=[n_{\rm
cl,x},x_{\rm in}]=\nonumber\\
&=&[n_{\rm cl,p},p_{\rm in}]=0.
\end{eqnarray}
Substituting Eqs.~(\ref{output}) and (\ref{classical}) into the
commutation rules
$[x_{\rm out},p_{\rm out}]=2i$ and (\ref{clcommutators}) one finds
using the
latter commutation rules (\ref{nxpcommutators1}) and
(\ref{nxpcommutators2}) that
the noise operators $n_{\rm{out,x}},n_{\rm{out,p}},n_{\rm{cl,x}}$ and
$n_{\rm{cl,p}}$
must satisfy
\begin{eqnarray}\label{commutators}
[n_{\rm{out,x}},n_{\rm{out,p}}]&=&2i\frac{(1-g^2)}{g^2},\quad
[n_{\rm{cl,x}},n_{\rm{cl,p}}]=-2i,\nonumber\\
\lbrack
n_{\rm{out,p}},n_{\rm{cl,x}}\rbrack&=&[n_{\rm{cl,p}},n_{\rm{out,x}}]=2i,\nonumber\\
\lbrack
n_{\rm{cl,x}},n_{\rm{out,x}}\rbrack&=&[n_{\rm{cl,p}},n_{\rm{out,p}}]=0.
\end{eqnarray}
The noise operators represent the noise by which the outcomes of the
homodyne detections
of the variables $x_{\rm cl}$ and $p_{\rm cl}$ as well as the output state are
contaminated.
The commutation rules (\ref{commutators}) and the Heisenberg uncertainty relations
then impose fundamental bounds on the noises that have to be satisfied by
any Gaussian operation. Since we are interested in partial measurements on
coherent states, it is convenient to quantify the two noises by the following sums:
\begin{equation}\label{sums}
\nu_{\rm out}\equiv\frac{\langle n_{\rm out,x}^{2}\rangle+
\langle n_{\rm out,p}^{2}\rangle}{2},\quad
\nu_{\rm cl}\equiv\frac{\langle n_{\rm cl,x}^{2}\rangle+
\langle n_{\rm cl,p}^{2}\rangle}{2},
\end{equation}
for which the respective bounds read as
\begin{equation}\label{uncertainty}
\nu_{\rm cl}\geq1,\quad \nu_{\rm out}\geq\frac{|1-g^{2}|}{g^2},\quad
\nu_{\rm cl}\nu_{\rm out}\geq 1.
\end{equation}
The use of noises (\ref{sums}) is advantageous since they are a
simple function of the added noises $\langle n_{\rm
out,x}^{2}\rangle$, $\langle n_{\rm out,p}^{2}\rangle$, $\langle
n_{\rm cl,x}^{2}\rangle$ and $\langle n_{\rm cl,p}^{2}\rangle$
that can be directly measured experimentally. We shall see that the operations that for a
given $\nu_{\mathrm{cl}}$ and $g$ minimize $\nu_{\mathrm{out}}$
 add noise symmetrically to the $x$ and $p$
quadratures, which means that $\langle n_{\rm out,x}^{2}\rangle=
\langle n_{\rm out,p}^{2}\rangle$ and $\langle n_{\rm
cl,x}^{2}\rangle=\langle n_{\rm cl,p}^{2}\rangle$ holds. In this case the quantities
$\nu_{\rm out}$ and $\nu_{\rm cl}$ are exactly the noises added to
the input state quadratures and to the measurement outcomes, respectively.
This symmetry and isotropy is a natural feature of optimal partial measurement on
coherent states that exhibit the same variances for all quadrature components.
The interpretation of noises (\ref{sums}) is particularly simple for
symmetric operations with unity gain ($g=1$). In this case the
quantity $\nu_{\rm out}/2$ coincides with the mean number of
thermal photons added by the operation to the input state. The
interpretation of the quantity $\nu_{\rm cl}$ is a little bit more
involved. The classical measurement outcomes $\bar{x}_{\rm cl}$
and $\bar{p}_{\rm cl}$ obtained when measuring the variables
$x_{\rm cl}$ and $p_{\rm cl}$  can be used to prepare a
classical guess $|\alpha_{\rm cl}\rangle=|(\bar{x}_{\rm cl}+i\bar{p}_{\rm
cl})/2\rangle$ of the input coherent state
$|\alpha\rangle=|(x+ip)/2\rangle_{\rm in}$. By repeating this procedure many
times with the same input state we thus prepare on average a mixed
quantum state called the estimated state of the input state. Similarly
as in the previous case for the symmetric unity gain operation the
quantity $(\nu_{\rm cl}+1)/2$ equals to the mean number of thermal
photons in the estimated state.

\section{Quantum teleportation as a minimal disturbance measurement}\label{sec1b}
In the following we show that one of the most celebrated quantum information protocols - quantum teleportation - enables a minimal disturbance measurement in the sense of saturating the inequalities in (\ref{uncertainty}). Using teleportation as an example we arrive at a very useful equality defining the optimum trade-off. The optimality will then be rigorously proven in the following two sections. 

The protocol in question is the standard continuous variable teleportation scheme
\cite{Vaidman_94,Braunstein_98,Furusawa_98} operating in the non-unity
gain regime~\cite{Bowen_03}. An unknown state of an optical mode ``in''
described by the quadratures $x_{\rm in}$ and $p_{\rm in}$
is teleported by a sender Alice ($A$) to a receiver Bob ($B$). At the
beginning,
Alice and Bob share an entangled state of two other modes $A$ and $B$
produced
by the two-mode squeezing transformation of two vacuum states
\begin{eqnarray}\label{TMSV}
x_{A}&=&\cosh(r)x_{A}^{(0)}-\sinh(r)x_{B}^{(0)},\nonumber\\
p_{A}&=&\cosh(r)p_{A}^{(0)}+\sinh(r)p_{B}^{(0)},\nonumber\\
x_{B}&=&\cosh(r)x_{B}^{(0)}-\sinh(r)x_{A}^{(0)},\nonumber\\
p_{B}&=&\cosh(r)p_{B}^{(0)}+\sinh(r)p_{A}^{(0)},
\end{eqnarray}
where $x_{A}^{(0)}$, $p_{A}^{(0)}$, $x_{B}^{(0)}$ and $p_{B}^{(0)}$
denote the vacuum quadratures
of modes $A$ and $B$ and $r$ is the squeezing parameter. Alice then
mixes the input mode with mode $A$ on a
balanced beam splitter and performs homodyne detection of the
variables
$x_{1}=(x_{\rm in}+x_{A})/\sqrt{2}$ and $p_{2}=(p_{\rm
in}-p_{A})/\sqrt{2}$ at the
outputs of the beam splitter. She then communicates the measurement
outcomes
$\bar{x}_{1}$ and $\bar{p}_{2}$ via a classical channel to Bob who
displaces
his part of the shared state as $x_{B}\rightarrow x_{\rm
out}=x_{B}+g\sqrt{2}\bar{x}_{1}$ and
$p_{B}\rightarrow p_{\rm out}=p_{B}+g\sqrt{2}\bar{p}_{2}$, where $g>0$
stands for the
gain of the transformation from photocurrents to the output optical field.
At Bob's site we thus
have the output quadratures (\ref{output}), where
\begin{eqnarray}\label{outnoise}
n_{\rm out,x}=x_{A}+\frac{x_{B}}{g},\quad n_{\rm
out,p}=-p_{A}+\frac{p_{B}}{g}.
\end{eqnarray}
At Alice's location we have two commuting variables (\ref{classical})
obtained by rescaling of the variables $x_{1}$ and $p_{2}$ by the
factor of $\sqrt{2}$
and the operators of added noises $n_{\rm cl,x}$ and $n_{\rm cl,p}$
read as
\begin{eqnarray}\label{classnoise}
n_{\rm cl,x}=x_{A},\quad n_{\rm cl,p}=-p_{A}.
\end{eqnarray}
Substituting now from Eqs.~(\ref{outnoise}) and (\ref{classnoise})
the noise operators
$n_{\rm{out,x}},n_{\rm{out,p}},n_{\rm{cl,x}}$ and $n_{\rm{cl,p}}$ in
the commutation rules
(\ref{commutators}) one finds the operators in the non-unity gain
teleportation
indeed satisfy the commutation algebra (\ref{commutators}). Making use
of
Eqs.~(\ref{TMSV}), (\ref{outnoise}) and (\ref{classnoise}) one
obtains the
noises (\ref{sums}) for the non-unity gain teleportation in the form,
\begin{eqnarray}\label{ntel}
\nu_{\rm cl}=\cosh(2r),\,\,\, \nu_{\rm
out}=\frac{(1+g^{2})}{g^{2}}\cosh(2r)-\frac{2}{g}\sinh(2r).\nonumber\\
\end{eqnarray}
It holds that $\langle n_{\mathrm{out},x}^2\rangle=\langle
n_{\mathrm{out},p}^2\rangle=\nu_{\mathrm{out}}$
and $\langle n_{\mathrm{cl},x}^2\rangle=\langle n_{\mathrm{cl},p}^2\rangle=\nu_{\mathrm{cl}}$
hence the added noise is isotropic.
Eliminating now the parameter $r$ from the second equation (\ref{ntel}) using the
first one one finds the trade-off between the noises (\ref{sums}) in the non-unity gain
teleportation to be
\begin{eqnarray}\label{trade-off}
g^{2}\nu_{\rm{out}}=(1+g^{2})\nu_{\rm{cl}}-2g\sqrt{\nu_{\rm{cl}}^{2}-1}.
\end{eqnarray}
In the plane of the noises $\nu_{\rm cl}$ and $\nu_{\rm out}$ the
trade-off relation
determines a certain quadratic curve that turns out to be a fraction of
a
hyperbola whose exact shape depends on the gain $g$. By changing the
squeezing $r$ one can continuously move along the whole trade-off curve
from
one extreme point to the other one. In the first extreme point one has
$\nu_{\rm cl}=1$,
i.e. the first of inequalities (\ref{uncertainty}) is saturated, while
$\nu_{\rm out}=(1+g^{2})/g^{2}$ and the point is reached for $r=0$. In
the second
extreme point the noise $\nu_{\rm out}$ attains the minimal possible
value $\nu_{\rm out}=|1-g^{2}|/g^{2}$,
i.e. the second of inequalities (\ref{uncertainty}) is saturated,
whereas
$\nu_{\rm cl}=\left|\frac{1+g^{2}}{1-g^{2}}\right|$ and the point is
reached
for $g>1$ ($g<1$) by choosing $r$ such that $\coth r=g$ ($\tanh r=g$).

Based on the previous results we arrive at an important property of
Gaussian quantum
operations described by the transformation rules (\ref{output}) and
(\ref{classical}).
Namely, in the plane $(\nu_{\rm cl},\nu_{\rm out})$ the optimal
operations lie in
the rectangle defined by the inequalities
\begin{eqnarray}
1&\leq&\nu_{\rm
cl}\leq\left|\frac{1+g^2}{1-g^2}\right|,\label{inequality1}\\
\frac{|1-g^{2}|}{g^{2}}&\leq&\nu_{\rm
out}\leq\frac{1+g^2}{g^{2}}.\label{inequality2}
\end{eqnarray}
The left-hand sides of the inequalities follow from the commutation
rules (\ref{commutators})
and cannot be overcome by any operation. On the other hand, the
operations that violate either
of the right-hand sides of the inequalities add too much noise and
therefore they are suboptimal.
This can be shown as follows. Consider
a quantum operation for which $\nu_{\rm out}'>(1+g^{2})/g^{2}$
($\nu_{\rm cl}''>\left|\frac{1+g^2}{1-g^2}\right|$).
The inequalities (\ref{uncertainty}) then reveal that at most 
$\nu_{\rm cl}'=1$ ($\nu_{\rm out}''=|1-g^{2}|/g^{2}$). Then, however,
we have
a better quantum operation given by the teleportation operating in the
first
(second) extreme point for which $\nu_{\rm cl}=1$ ($\nu_{\rm
out}=|1-g^{2}|/g^{2}$)
but simultaneously $\nu_{\rm out}=(1+g^{2})/g^{2}<\nu_{\rm out}'$
($\nu_{\rm cl}=\left|\frac{1+g^{2}}{1-g^{2}}\right|<\nu_{\rm cl}''$).

\begin{figure}
\centerline{\psfig{width=9.0cm,angle=0,file=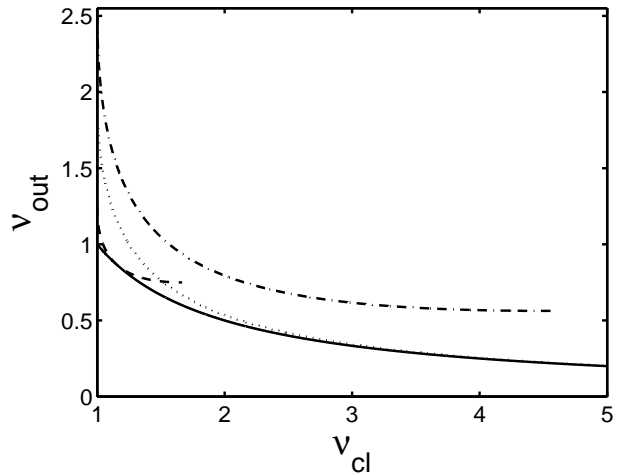}}
\caption{Optimal trade-off between the output noise $\nu_{\rm out}$ and
the noise in
measurement outcomes $\nu_{\rm cl}$ for a single-mode Gaussian
operation with optimal
gain $g=\nu_{\rm cl}/\sqrt{\nu_{\rm cl}^{2}-1}$ (solid curve),
amplifying operation with
$g=2$ (dashed curve), unity gain operation with $g=1$ (dotted curve)
and attenuating operation with $g=0.8$ (dash-dotted curve). See text
for details.}
\label{fig1}
\end{figure}

The formulas (\ref{trade-off}), (\ref{inequality1}) and
(\ref{inequality2}) are one of
the main theoretical results of the present paper. This is because as
we will show in the following
section the trade-off (\ref{trade-off}) is optimal on the set of all
Gaussian operations
described by Eqs.~(\ref{output}), (\ref{classical}) and
(\ref{commutators}). The trade-off
is depicted for several values of the gain $g$ in Fig.~\ref{fig1}.

Before going to the proof of optimality we can answer another
important question based on the trade-off (\ref{trade-off}). Up to
now we considered Gaussian quantum operations with a fixed gain
$g$. Provided that the trade-off (\ref{trade-off}) is optimal its
right-hand side then gives us (after division by $g^{2}$) the
least possible noise $\nu_{\rm out}$ that can be attained for a
given value of noise $\nu_{\rm cl}$ by any such operation. The
fundamental question that can be risen in this context is that if
the gain $g$ of operation can be adjusted freely what is its
optimal value $g_{\rm opt}$ that gives for a given value of the
noise $\nu_{\rm cl}$ the least possible value of the noise
$\nu_{\rm out}$. The task was already solved by Ralph
\cite{Ralph_00.pra} who showed that in the non-unity gain
teleportation one can adjust for a given value of the noise
$\nu_{\rm cl}$ the gain such that the third of inequalities
(\ref{uncertainty}) is saturated and therefore such
teleportation protocol realizes the sought optimal operation. The
trade-off relation (\ref{trade-off}) contains Ralph's result as a
particular instance and can be used to rederive it: Expressing
$\nu_{\rm out}$ as a function of $g$ and $\nu_{\rm cl}$ using
Eq.~(\ref{trade-off}) and minimizing it with respect to $g$ one
finds the optimal gain for $\nu_{\rm cl}\ne 1$ to be $g_{\rm
opt}=\nu_{\rm cl}/\sqrt{\nu_{\rm cl}^{2}-1}$ that gives $\nu_{\rm
out}=1/\nu_{\rm cl}$ and thus the fundamental quantum mechanical
limit given by the third of inequalities (\ref{uncertainty}) is
indeed saturated. For $\nu_{\rm cl}=1$ the optimal gain is
infinitely large ($g_{\rm opt}=\infty$) for which one has
$\nu_{\rm out}=1$. In this case, all the inequalities
(\ref{uncertainty}) are saturated simultaneously but the operation
achieving this regime is unphysical.

\section{Proofs of optimality}\label{sec_2}

In this section we prove the optimality of the inequalitites derived above using two different methods. The optimization task we want to solve can be generally formulated
as follows: Find
a Gaussian operation described by Eqs.~(\ref{output}),
(\ref{classical}) and (\ref{commutators})
that for a given gain $g$ and a given amount of added noise in the
measurement outcomes adds
the least possible amount of noise into the input state.
The optimal operation will in general depend on
the quantities used to quantify the two noises. Here we are interested
in optimal operations
that add noise symmetrically into the amplitude and phase quadrature,
i.e. for which
$\langle n_{\rm out,x}^{2}\rangle=\langle n_{\rm out,p}^{2}\rangle$ and
$\langle n_{\rm cl,x}^{2}\rangle=\langle n_{\rm cl,p}^{2}\rangle$. As
we will show below, this requirement is satisfied if we take the
sums (\ref{sums}) of the variances of the
noise operators $n_{\rm{cl,x}},n_{\rm{cl,p}},n_{\rm{out,x}}$ and
$n_{\rm{out,p}}$ to
quantify the noise in the measurement outcomes and the noise added into
the input state, respectively.

This is a consequence of the fact that for any Gaussian operation that
is asymmetric in $x$ and $p$
variables, i.e. for which $\langle n_{\rm out,x}^{2}\rangle\ne\langle
n_{\rm out,p}^{2}\rangle$
and $\langle n_{\rm cl,x}^{2}\rangle\ne\langle n_{\rm
cl,p}^{2}\rangle$, there is always a symmetric
Gaussian operation giving the same values of $\nu_{\rm cl}$ and
$\nu_{\rm out}$. This statement can
be proved in the following way. Suppose we have the asymmetric
operation described by the formulas
\begin{eqnarray}\label{operation1}
x_{\rm out}&=&g(x_{\rm in}+n_{\rm{out,x}}),\quad x_{\rm cl}=x_{\rm
in}+n_{\rm{cl,x}},\nonumber\\
p_{\rm out}&=&g(p_{\rm in}+n_{\rm{out,p}}),\quad p_{\rm cl}=p_{\rm
in}+n_{\rm{cl,p}}.
\end{eqnarray}
Assume that, in addition, we have at our disposal another asymmetric
operation that is obtained
from the previous one by placing it in between one phase shifter at the input and two
phase-shifters at the outputs.
The first phase shifter interchanges the input quadratures as follows,
$x_{\rm in}\rightarrow -p_{\rm in}$ and $p_{\rm in}\rightarrow x_{\rm in}$,
and the two phase shifters on the output modes perform the inverse transformation
$x_{i}\rightarrow p_{i}$ and $p_{\rm i}\rightarrow -x_{\rm i}$, $i={\rm out,cl}$.
Taking all the above transformation rules together the entire operation is described
by the following rules:
\begin{eqnarray}\label{operation2}
x_{\rm out}&=&g(x_{\rm in}+n_{\rm{out,p}}'),\quad x_{\rm cl}=x_{\rm
in}+n_{\rm{cl,p}}',\nonumber\\
p_{\rm out}&=&g(p_{\rm in}-n_{\rm{out,x}}'),\quad p_{\rm cl}=p_{\rm
in}-n_{\rm{cl,x}}'.
\end{eqnarray}
The prime was used merely to express that the noise operators in
Eq.~(\ref{operation2}) are completely independent on and therefore
completely uncorrelated with the unprimed noise operators in
Eq.~(\ref{operation1}). The variances of the primed and unprimed
noise operators are, however, identical, $\langle n_{\rm
out,i}^{2}\rangle=\langle (n_{\rm out,i}')^{2}\rangle$ and
$\langle {n_{\rm cl,i}}^{2}\rangle=\langle (n_{\rm
cl,i}')^{2}\rangle$, $i=x,p$. The desired symmetric operation can
then be constructed from the operations (\ref{operation1}) and
(\ref{operation2}) by placing them into two arms of a balanced
Mach-Zehnder interferometer. At the first balanced beam splitter
of the interferometer the input quadratures are mixed with the
quadratures $x_{0}$ and $p_{0}$ of an auxiliary mode $0$ as
$x_{\rm in}'=(x_{\rm in}-x_{0})/\sqrt{2}$, $p_{\rm in}'=(p_{\rm
in}-p_{0})/\sqrt{2}$ and $x_{0}'=(x_{\rm in}+x_{0})/\sqrt{2}$,
$p_{0}'=(p_{\rm in}+p_{0})/\sqrt{2}$. The quadratures $x_{\rm
in}'$, $p_{\rm in}'$ and $x_{0}'$, $p_{0}'$ are then used as
inputs into the operation (\ref{operation1}) and
(\ref{operation2}), respectively. The quadratures at outputs of
the operations $x_{\rm in}''$, $p_{\rm in}''$, $x_{0}''$ and
$p_{0}''$ are finally superimposed on the second balanced beam
splitter of the interferometer at one outcome of which one has
\begin{eqnarray}\label{symout}
x_{\rm out}&=&\frac{x_{\rm in}''+x_{0}''}{\sqrt{2}}=
g(x_{\rm in}+\tilde{n}_{\rm{out,x}}),\nonumber\\
p_{\rm out}&=&\frac{p_{\rm in}''+p_{0}''}{\sqrt{2}}=
g(p_{\rm in}+\tilde{n}_{\rm{out,p}}),
\end{eqnarray}
where
$\tilde{n}_{\rm{out,x}}=({n}_{\rm{out,x}}+{n}_{\rm{out,p}}')/\sqrt{2}$ and
$\tilde{n}_{\rm{out,p}}=({n}_{\rm{out,p}}-{n}_{\rm{out,x}}')/\sqrt{2}$.
Further, two pairs of the commuting variables $x_{\rm cl,in}$ $p_{\rm
cl,in}$
and $x_{{\rm cl},0}$ $p_{{\rm cl},0}$ representing the output of the
operation
(\ref{operation1}) on mode ``in'' of the interferometer and the
operation
(\ref{operation2}) on mode $0$, respectively, give after averaging a
new pair of
commuting variables
\begin{eqnarray}\label{symclass}
x_{\rm cl}&=&\frac{x_{\rm cl,in}+x_{{\rm cl},0}}{\sqrt{2}}=
x_{\rm in}+\tilde{n}_{\rm{cl,x}},\nonumber\\
p_{\rm cl}&=&\frac{p_{\rm cl,in}+p_{{\rm cl},0}}{\sqrt{2}}=
p_{\rm in}+\tilde{n}_{\rm{cl,p}},
\end{eqnarray}
where
$\tilde{n}_{\rm{cl,x}}=({n}_{\rm{cl,x}}+{n}_{\rm{cl,p}}')/\sqrt{2}$
and
$\tilde{n}_{\rm{cl,p}}=({n}_{\rm{cl,p}}-{n}_{\rm{cl,x}}')/\sqrt{2}$.
As the primed and the unprimed noise operators are uncorrelated
one immediately finds that $\langle \tilde{n}_{\rm
out,x}^{2}\rangle= \langle \tilde{n}_{\rm out,p}^{2}\rangle$ as
well as $\langle \tilde{n}_{\rm cl,x}^{2}\rangle= \langle
\tilde{n}_{\rm cl,p}^{2}\rangle$ and therefore the new operation
described by Eqs.~(\ref{symout}) and (\ref{symclass}) is symmetric
with respect to $x$ and $p$. Moreover, calculating the noises
(\ref{sums}) for the new operation yields $\tilde{\nu}_{\rm
out}=\langle \tilde{n}_{\rm out,x}^{2}\rangle=\nu_{\rm out}$ and
$\tilde{\nu}_{\rm cl}=\langle \tilde{n}_{\rm
cl,x}^{2}\rangle=\nu_{\rm cl}$ which completes the proof.
\subsection{Proof I}

For the sake of simplicity of mathematical formulas occurring in the
proofs of optimality of the trade-off (\ref{trade-off}) we will work
with rescaled operators of added noises
\begin{equation}\label{moperators}
m_{\rm{out,x}}\equiv g n_{\rm{out,x}},\quad
m_{\rm{out,p}}\equiv g n_{\rm{out,p}}.
\end{equation}
Using these new operators one can write $\nu_{\rm out}=\sigma_{\rm
out}/g^{2}$, where
\begin{equation}
\sigma_{\rm out}\equiv\frac{\langle m_{\rm out,x}^{2}\rangle+
\langle m_{\rm out,p}^{2}\rangle}{2},
\end{equation}
and the trade-off (\ref{trade-off}) whose optimality is to be
proved then reads
\begin{eqnarray}\label{trade-off2}
\sigma_{\rm{out}}=(1+g^{2})\nu_{\rm{cl}}-2g\sqrt{\nu_{\rm{cl}}^{2}-1}.
\end{eqnarray}
It is convenient to introduce the
column vector
$\tau=(n_{\rm{cl,x}},m_{\rm{out,x}},n_{\rm{cl,p}},m_{\rm{out,p}})^{\rm T}$.
In this notation all the commutators (\ref{commutators}) can be
rewritten in the compact form
$[\tau_{i},\tau_{j}]=2i\Gamma_{ij}$, where
\begin{eqnarray}\label{omega}
\Gamma=\left(\begin{array}{cc}
0 & -G \\
G & 0 \\
\end{array}\right),\quad  G=\left(\begin{array}{cc}
1 & g \\
g & -(1-g^2) \\
\end{array}\right).\
\end{eqnarray}
Since the gains of considered Gaussian operations are fixed the first
moments of the noise operators
vanish, i.e. $\langle\tau\rangle=0$, where the symbol
$\langle\,\,\,\rangle$
denotes averaging over the input state $\rho_{\rm aux}$ of the
auxiliary modes. Consequently, the studied
operations are completely characterized by the $4\times 4$ real
symmetric noise
matrix $N$ with elements $N_{ij}=\langle\{\tau_{i},\tau_{j}\}\rangle$,
where $\{A,B\}\equiv(AB+BA)/2$. The
commutation rules (\ref{commutators}) then impose a specific
uncertainty principle
on the noise matrix $N$ that reads
\begin{equation}\label{constraint}
    N+i\Gamma\geq 0.
\end{equation}

Now we want to find such of the considered quantum operations that
gives for a given noise
$\nu_{\rm cl}$ minimum possible noise $\sigma_{\rm out}$. This task can
be
equivalently reformulated as follows:
\begin{eqnarray}\label{SDP}
\mathop{\mathrm{minimize}}_{N}\quad f(N)=a\,\nu_{\rm cl}+b\,\sigma_{\rm
out},
\end{eqnarray}
under the constraint (\ref{constraint}). The coefficients $a,b\geq0$
(except for the case $a=b=0$)
control the ratio between the noise in the measurement outcomes and in
the output state. The
optimization task (\ref{SDP}) is a typical example of the so-called
semidefinite programme
(SDP) \cite{Vandenberghe_96}. Recently, also other important problems
in quantum information
theory were formulated and solved as
semidefinite programmes ranging from separability criteria
\cite{Doherty_02,Hyllus_06} and
optimization of completely positive maps \cite{Audenaert_02} to
optimization of teleportation
with a mixed entangled state \cite{Verstraete_03} or finding optimal
POVMs for quantum state
discrimination \cite{Jezek_02,Eldar_03}.

The SDPs are generally difficult to solve analytically and we are often
forced to use numerical
methods. However, in the case of the problem (\ref{SDP}) we are able to
find the solution
analytically. This can be done in two steps following the standard
strategy employed, for instance,
in \cite{Audenaert_02,Fiurasek_06}. In the first step we guess the
analytical form of the solution of the
problem (\ref{SDP}) while in the second step we prove its optimality.
The first step has been
already done in the previous section where we surmised the solution of
the problem (\ref{SDP})
to be given by the non-unity gain teleportation described by
Eqs.~(\ref{TMSV})-(\ref{classnoise}).
Calculating the operators (\ref{moperators}) using
Eqs.~(\ref{outnoise}) and substituting
them together with the operators (\ref{classnoise}) into the definition
of the noise
matrix $N$ we arrive at the noise matrix for the teleportation in the
form:
\begin{equation}\label{Ntel}
N_{\rm tel}=A\oplus A,
\end{equation}
where $A$ is the symmetric $2\times 2$ matrix with elements
\begin{eqnarray}\label{A}
A_{11}&=&\nu_{\rm cl},\quad A_{12}=A_{21}=g \nu_{\rm cl}-\sqrt{\nu_{\rm
cl}^{2}-1},\nonumber\\
A_{22}&=&(1+g^{2}) \nu_{\rm cl}-2g\sqrt{\nu_{\rm cl}^{2}-1},
\end{eqnarray}
where $\nu_{\rm cl}=\cosh(2r)$. Since the matrix (\ref{Ntel}) is
manifestly invariant under
the exchange of subscripts $x$ and $p$ the noise is added symmetrically
into the amplitude and phase quadrature as required and the non-unity gain
teleportation is indeed a good candidate for the optimal operation.

In order to prove optimality of the matrix (\ref{Ntel}) we can proceed
along
the lines of the proof of optimality of multicopy asymmetric cloning of
coherent states
\cite{Fiurasek_06}. The proof relies on finding a certain Hermitean
positive semidefinite
$4\times4$ matrix $Z$ that satisfies for any admissible matrix $N$ the
condition
$\mathrm{Tr}(ZN)=f(N)$. From the condition $Z\geq0$ and the constraint
$N+i\Gamma\geq0$ then
immediately follows a lower bound on the functional $f(N)$ that is to be minimized,
$f(N)=\mathrm{Tr}(ZN)\geq-i\mathrm{Tr}(Z\Gamma)$. If, in addition, $Z$
satisfies the condition
\begin{equation}\label{CS}
Z(N_{\rm tel}+i\Gamma)=0,
\end{equation}
the lower bound is saturated by the matrix (\ref{Ntel}) and
therefore the
corresponding quantum operation is optimal.

The matrix $Z$ we are looking for can be taken in the block form
\cite{Fiurasek_06}
\begin{eqnarray}\label{Z}
Z=\frac{1}{2}\left(\begin{array}{cc}
P & iQ \\
-iQ & P \\
\end{array}\right),
\end{eqnarray}
where $P$ and $Q$ are real symmetric $2\times2$ matrices. The condition
(\ref{CS}) gives rise
to the following set of equations for the matrices $P$ and $Q$
\begin{equation}\label{PQequations}
Q\Gamma=PA,\quad QA=P\Gamma.
\end{equation}
The matrix $P$ is determined solely by the condition
$\mathrm{Tr}(ZN)=f(N)$ that gives $P=\mbox{diag}(a,b)$.
This is because Eqs.~(\ref{PQequations}) do not impose any further
restriction on $P$ as they provide
the equation $P(A\Gamma^{-1}A-\Gamma)=0$ for the matrix $P$ that is
satisfied by any $P$ due to the equality
$A\Gamma^{-1}A=\Gamma$. Having the matrix $P$ in hands one can now
substitute it into the equation
$Q=PA\Gamma^{-1}$ derived from the first of Eqs.~(\ref{PQequations})
that leads to the matrix
$Q$ in the form:
\small
\begin{eqnarray}\label{Q}
Q=
\left(\begin{array}{cc}
a\left(\nu_{\rm cl}-g\sqrt{\nu_{\rm cl}^{2}-1}\right) & a\sqrt{\nu_{\rm
cl}^{2}-1}\\
b\left[2g\nu_{\rm cl}-(1+g^{2})\sqrt{\nu_{\rm cl}^{2}-1}\right] &
b\left(g\sqrt{\nu_{\rm cl}^{2}-1}-\nu_{\rm cl}\right)\\
\end{array}\right).\nonumber\\
\end{eqnarray}
\normalsize
For our guess (\ref{Ntel}) the coefficients $a$ and $b$ are not
independent but instead they
are tied together by a specific relation that can be calculated by
minimizing the functional
$f(N)$ under the constraint (\ref{trade-off}). Using the standard
method of Lagrange multipliers
one then finds the relation to be
\begin{equation}\label{abrelation}
a\sqrt{\nu_{\rm cl}^{2}-1}=b\left[2g\nu_{\rm cl}-(1+g^2)\sqrt{\nu_{\rm
cl}^{2}-1}\right],
\end{equation}
which reveals that the matrix $Q$ is indeed symmetric.
It remains to check the positive semidefiniteness of the matrix
(\ref{Z}). Since we require $a,b\geq0$
(except for the case $a=b=0$) the expression in the square brackets on
the right hand side of Eq.~(\ref{abrelation})
must be nonnegative. This condition is not satisfied exactly by those
operations which violate the inequality $\nu_{\rm
cl}\leq\left|\frac{1+g^{2}}{1-g^{2}}\right|$.
Since, however, these operations have been already
ruled out from our considerations as being suboptimal it is sufficient
to restrict ourselves to operations
satisfying this inequality. 
For these operations three
different cases must be distinguished in dependence on the value of the
noise $\nu_{\rm cl}$.\\
$1)$ If $\nu_{\rm cl}=1$ then Eq.~(\ref{abrelation}) implies $b=0$
whence $P=Q=\mbox{diag}(a,0)$.
The eigenvalues of the matrix $Z$ then read as $\alpha_{1,2,3}=0$ and
$\alpha_{4}=a>0$ and therefore
$Z\geq0$.\\
$2)$ For $\nu_{\rm cl}=\left|\frac{1+g^{2}}{1-g^{2}}\right|$ one finds
$a=0$ using Eq.~(\ref{abrelation})
that gives $P=\mbox{diag}(0,b)$ and $Q=\mbox{diag}(0,\pm b)$ where the
upper (lower) sign holds for
$g>1$ ($g<1$). One can again directly calculate the eigenvalues of the
matrix $Z$ in the form
$\beta_{1,2,3}=0$ and $\beta_{4}=b>0$ and therefore $Z\geq0$.\\
$3)$ In the intermediate case when $1<\nu_{\rm
cl}<\left|\frac{1+g^{2}}{1-g^{2}}\right|$ one has $a>0$ and
simultaneously $b>0$. This allows one to introduce the matrix
$V=\sqrt{2}\mbox{diag}(P^{-\frac{1}{2}},P^{-\frac{1}{2}})$
and to transform the matrix $Z$ as
\begin{eqnarray}\label{V}
Z_{1}=V^{\dag}ZV=\left(\begin{array}{cc}
I & iP^{-\frac{1}{2}}QP^{-\frac{1}{2}} \\
-iP^{-\frac{1}{2}}QP^{-\frac{1}{2}} & I \\
\end{array}\right),
\end{eqnarray}
where $I$ is the $2\times2$ identity matrix. The specific feature of
the transformation
(\ref{V}) is that if $Z_{1}$ is positive semidefinite then also $Z$ is
positive semidefinite
and it is thus sufficient to prove the positive semidefiniteness of the
matrix $Z_{1}$. Performing the similarity transformation
$Z_{2}=UZ_{1}U^{\dag}$ where
\begin{eqnarray}
U=\frac{1}{\sqrt{2}}\left(\begin{array}{cc}
I & iI \\
iI & I \\
\end{array}\right),
\end{eqnarray}
the matrix $Z_{1}$ is brought into the block diagonal matrix
$Z_{2}=\mbox{diag}(I+P^{-\frac{1}{2}}QP^{-\frac{1}{2}},I-P^{-\frac{1}{2}}QP^{-\frac{1}{2}})$
whose eigenvalues are easy to find in the form $\gamma_{1,2}=0$ and
$\gamma_{3,4}=2$.
Consequently, $Z_{2}\geq0$ and therefore also $Z\geq0$ which completes
the proof.
\subsection{Proof II}

There is an alternative way of proving the optimality of the noise
matrix (\ref{Ntel}). The
proof relies on the mapping of the noise operators $m_{{\rm out},x}$,
$m_{{\rm out},p}$,
$n_{\rm{cl,x}}$ and  $n_{\rm{cl,p}}$ onto the quadratures in the
non-unity gain teleportation.
The search for the optimal noise matrix then boils down to searching 
a suitable two-mode state
shared in the non-unity gain teleportation.

Suppose we have a noise matrix $N$, i.e. a real symmetric $4\times4$
matrix satisfying the uncertainty
principle $N+i\Gamma\geq 0$. Assume in addition, there is a real
regular $4\times4$ matrix $M$ satisfying the
condition
\begin{equation}\label{mappingmatrix}
\Gamma=M\Omega M^{\rm T},
\end{equation}
which means that $M$ realizes mapping between the commutation rules for
noise operators
$[\tau_{i},\tau_{j}]=2i\Gamma_{ij}$ and the standard canonical
commutation rules
$[\xi_{i},\xi_{j}]=2i\Omega_{ij}$, where
$\xi=(x_{A},p_{A},x_{B},p_{B})^{\rm T}$ is
the vector of quadratures and
\begin{eqnarray}\label{Omega}
\Omega=J\oplus J,\quad
J=\left(\begin{array}{cc}
0 & 1 \\
-1 & 0 \\
\end{array}\right)
\end{eqnarray}
is the standard symplectic matrix. Provided that such a matrix $M$
exists we can associate with
any admissible noise matrix $N$ a certain real symmetric $4\times4$
matrix
\begin{eqnarray}\label{variancematrix}
V_{AB}=M^{-1}N\left(M^{\rm T}\right)^{-1},
\end{eqnarray}
that can be shown to satisfy the standard Heisenberg uncertainty
principle
$V_{AB}+i\Omega\geq0$ and therefore to be a covariance matrix of a
two-mode state.
This can be shown as follows. Expressing the left hand side of the
Heisenberg uncertainty
principle using the formuls (\ref{mappingmatrix}) and
(\ref{variancematrix}) one finds
$V_{AB}+i\Omega=M^{-1}(N+i\Gamma)(M^{T})^{-1}$. Taking now the spectral
decomposition
$N+i\Gamma=\sum_{i}\mu_{i}|\mu_{i}\rangle\langle\mu_{i}|$, where
$\mu_{i}\geq0$ are
eigenvalues and $|\mu_{i}\rangle$ are corresponding eigenvectors of the
matrix $N+i\Gamma$, one finds that
$\langle\psi|V_{AB}+i\Omega|\psi\rangle=\sum_{i}\mu_{i}|\langle\psi|M^{-1}|\mu_{i}\rangle|^{2}\geq0$
for any vector $|\psi\rangle$ and therefore $V_{AB}+i\Omega$ is indeed
positive semidefinite whence $V_{AB}$
is a two-mode covariance matrix. Since $V_{AB}$ is a covariance matrix its
elements can be written as
$(V_{AB})_{ij}=\mbox{Tr}\left[\rho_{AB}\{\xi_{i},\xi_{j}\}\right]$,
$i,j=1,\ldots,4$,
where $\xi_{i}$, $i=1,\ldots,4$ are components of the vector $\xi$ of
standard
quadratures and $\rho_{AB}$ is a state of two modes $A$ and $B$. A
natural realization
of the vector of quadratures $\xi$ is by the linear relation
\begin{eqnarray}\label{mapping}
\xi=M^{-1}\tau.
\end{eqnarray}
In this case the state $\rho_{AB}$ coincides with the state $\rho_{\rm
aux}$ over
which the averaging in the definition of the noise matrix $N$ is
performed.

It remains to show that a regular matrix $M$ satisfying the condition
(\ref{mappingmatrix})
exists. Non-unity gain teleportation provides such a matrix that, in
addition, proves to be
suitable for minimization of the functional (\ref{SDP}). By writing
Eqs.~(\ref{outnoise}) and
(\ref{classnoise}) in the matrix form (\ref{mapping}) one finds the
matrix $M$ in the non-unity
gain teleportation to be a regular matrix of the form:
\begin{eqnarray}\label{M}
M=\left(\begin{array}{cccc}
1 & 0 & 0 & 0 \\
g & 0 & 1 & 0 \\
0 & -1 & 0 & 0 \\
0 & -g & 0 & 1 \\
\end{array}\right).
\end{eqnarray}
The problem of minimization of the functional (\ref{SDP}) over the noise
matrices $N$ is then
transformed into the problem of finding a two-mode state $\rho_{AB}$
(with covariance matrix $V_{AB}$)
shared in the non-unity gain teleportation that minimizes the
functional (\ref{SDP}). Substituting
from Eqs.~(\ref{outnoise}) and (\ref{classnoise}) into Eq.~(\ref{SDP})
using the definitions
(\ref{sums}) one can express the functional (\ref{SDP}) as the trace
$f(N)=\mbox{Tr}(\rho_{AB}O)$, where
\begin{eqnarray}\label{O}
O&=&\frac{1}{2}\left[a\left(g^{2}x_{A}^{2}+2g\{x_{A},x_{B}\}+x_{B}^{2}+g^{2}p_{A}^{2}\right.\right.\nonumber\\
&&\left.\left.-2g\{p_{A},p_{B}\}+p_{B}^{2}\right)+b\left(x_{A}^{2}+p_{A}^{2}\right)\right].
\end{eqnarray}
The operator $O$ is lower bounded by
$O\geq\mbox{min}[\mbox{eig}(O)]\openone$ which
implies that the functional $f(N)$ is lower bounded by
$f(N)\geq\mbox{min}[\mbox{eig}(O)]$. This lower
bound is saturated if the state $\rho_{AB}$ is an eigenstate
of the operator (\ref{O})
corresponding to its lowest eigenvalue. The operator (\ref{O}) can be
diagonalized by the two-mode squeezing
transformation $S$ described in the Heisenberg picture by
Eq.~(\ref{TMSV}).
Choosing the squeezing parameter as
\begin{equation}\label{diagsqueezing}
\tanh{\left(2r\right)}=\frac{2ga}{a\left(g^{2}+1\right)+b}
\end{equation}
the operator (\ref{O}) is diagonalized to the form
\begin{equation}\label{Odiagonal}
O'=SOS^{\dag}=2(xn_{A}+yn_{B})+x+y,
\end{equation}
where $n_{i}=(x_{i}^{2}+p_{i}^{2})/4-1/2$, $i=A,B$ are standard photon
number operators and $x,y\geq0$.
Inserting the formula (\ref{Odiagonal}) into the expression for
$f(N)=\mbox{Tr}(\rho_{AB}O)$ one obtains
that $f(N)=\mbox{Tr}(S\rho_{AB}S^{\dag}O')$ which is obviously
minimized if
$S\rho_{AB,{\rm opt}}S^{\dag}=|00\rangle\langle00|$, where $|00\rangle$
is the vacuum state. Hence, the optimal
state $\rho_{AB,{\rm opt}}$ is the two-mode squeezed vacuum state with
the squeezing parameter given by the
formula (\ref{diagsqueezing}). Thus, we arrived in a different way at
the conclusion that the non-unity
gain teleportation with shared two-mode squeezed vacuum state with a
properly chosen squeezing represents
optimal Gaussian quantum operation that for a given gain $g$ and noise
$\nu_{\rm cl}$ introduces the least possible noise $\nu_{\rm out}$.

\section{Linear optics scheme}\label{sec_3}

\begin{figure}[!h]
\centering \includegraphics[width=7cm]{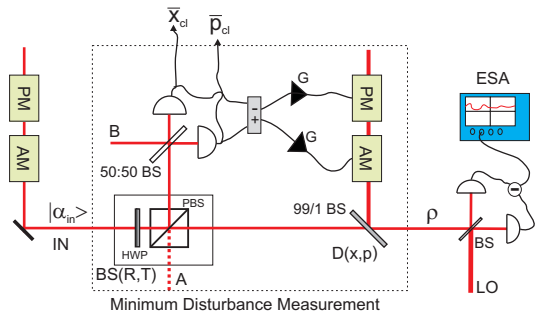} \caption{The
Experimental Scheme. AM: Amplitude Modulator; PM: Phase Modulator;
$|\alpha \rangle$: The incoming coherent state to be measured;
HWP: Half Wave Plate; PBS: Polarizing Beam Splitter Cube; BS: Beam
Splitter; D(x,p): Displacement operation; B: Auxiliary beam; BS(R,T): Variable beam
splitter with reflectivity $R$ and transmissivity $T$ for intensity; A: Vacuum mode;
G: electronic gains; $x_{\rm cl}, p_{\rm cl}$: classical measurement
outcomes; $\rho$: output state; LO: Local Oscillator Beam;
ESA: Electronic Spectrum Analyzer.} \label{scheme}
\end{figure}

Non-unity gain teleportation is not the only scheme that 
saturates the optimal trade-off (\ref{trade-off}). As shown in Ref.~\cite{Andersen_06} there are at least two other schemes that can accomplish a minimal disturbance measurement. One other strategy is to use optimal 1$\rightarrow$2 Gaussian cloning followed by a joint measurement between one of the clones and the anti-clone. However, a much simpler strategy which will be investigated in the following achieves the optimal bound using only linear optics, homodyne detection and feed-forward. The setup is depicted in Fig.~\ref{scheme}.
In this scheme the input mode ``in'' is
mixed with an auxiliary vacuum mode $A$ on an unbalanced beam
splitter with amplitude reflectivity $\sqrt{R}$ and transmissivity
$\sqrt{T}$ ($R+T=1$). The reflected mode
``$\mbox{in}$'' and the transmitted mode $A$ are described by the
following quadratures
\begin{eqnarray}\label{unbalancedBS}
x_{\rm in}'&=&\sqrt{R}x_{\rm in}+\sqrt{T}x_{A}^{(0)},\quad p_{\rm
in}'=\sqrt{R}p_{\rm in}+\sqrt{T}p_{A}^{(0)},\nonumber\\
x_{A}'&=&\sqrt{T}x_{\rm in}-\sqrt{R}x_{A}^{(0)},\quad
p_{A}'=\sqrt{T}p_{\rm in}-\sqrt{R}p_{A}^{(0)}.\nonumber\\
\end{eqnarray}
The mode ``$\mbox{in}$'' is then superimposed with another vacuum mode
$B$ on a balanced beam
splitter and the quadratures $x_{1}\equiv x_{\rm in}''=(x_{\rm
in}'+x_{B}^{(0)})/\sqrt{2}$ and
$p_{2}\equiv p_{A}'=(p_{\rm in}'-p_{B}^{(0)})/\sqrt{2}$ are measured at
its outputs. After
rescaling the measured quadratures $x_{1}$ and $p_{2}$ by the factor
$\sqrt{2/R}$ we arrive at
the variables $x_{\rm cl}$ and $p_{\rm cl}$ in the form
(\ref{classical}), where
\begin{equation}\label{ncl}
n_{\rm cl,x}=\frac{\sqrt{T}x_{A}^{(0)}+x_{B}^{(0)}}{\sqrt{R}},\quad
n_{\rm cl,p}=\frac{\sqrt{T}p_{A}^{(0)}-p_{B}^{(0)}}{\sqrt{R}}.
\end{equation}
The outcomes of the measurement $\bar x_{1}$ and $\bar p_{2}$ are then
used to displace the
mode $A$ as $x_{A}'\rightarrow x_{\rm out}=x_{A}'+\sqrt{2}G\bar{x}_{1}$
and
$p_{A}'\rightarrow p_{\rm out}=p_{A}'+\sqrt{2}G\bar{p}_{2}$, where $G$
is the normalized
electronic gain. The output quadratures $x_{\rm out}$ and $p_{\rm out}$
are then of the form
(\ref{output}), where the gain reads as
\begin{equation}\label{gain}
g=\sqrt{T}+G\sqrt{R},
\end{equation}
and
\begin{eqnarray}\label{nout}
n_{\rm
out,x}&=&\frac{\left(G\sqrt{T}-\sqrt{R}\right)x_{A}^{(0)}+Gx_{B}^{(0)}}{g},\nonumber\\
\quad n_{\rm
out,p}&=&\frac{\left(G\sqrt{T}-\sqrt{R}\right)p_{A}^{(0)}-Gp_{B}^{(0)}}{g}.
\end{eqnarray}
Inserting now Eqs.~(\ref{ncl}) and (\ref{nout}) into the definitions
(\ref{sums}) one finds
\begin{eqnarray}\label{nufeed-forward}
\nu_{\rm cl}=\frac{1+T}{1-T},\quad \nu_{\rm
out}=\frac{\left(G\sqrt{T}-\sqrt{R}\right)^{2}+G^{2}}{g^{2}}.
\end{eqnarray}
Expressing $g^{2}\nu_{\rm out}$ using the second of equations
(\ref{nufeed-forward}), substituting in the obtained formula for
$G$ employing Eq.~(\ref{gain}) and making use of the formula $R+T=1$
and the first of equations (\ref{nufeed-forward}) we finally
confirm that the noises (\ref{nufeed-forward}) indeed satisfy the
optimal trade-off (\ref{trade-off}).

Equations (\ref{nufeed-forward}) immediately allow us to derive
the output noise $\nu_{\rm out}$ for the case with optimized gain.
Making use of the first of equations (\ref{nufeed-forward}) in the
formula $g_{\rm opt}=\nu_{\rm cl}/\sqrt{\nu_{\rm cl}^{2}-1}$ one
gets the optimal gain
\begin{equation}\label{gopt}
g_{\rm opt}=\frac{1+T}{2\sqrt{T}}.
\end{equation}
Substituting further the latter expression for $g_{\rm opt}$ into
Eq.~(\ref{gain}) one finds the electronic gain $G_{g_{\rm opt}}$
to be $G_{g_{\rm opt}}=\sqrt{1-T}/2\sqrt{T}$. Inserting now
Eq.~(\ref{gopt}) and the obtained expression for $G_{g_{\rm opt}}$
into the second of equations (\ref{nufeed-forward}) we finally
arrive at the output noise in the linear optics scheme with the
feed-forward in the form
\begin{eqnarray}\label{nuoutfeed-forward}
\nu_{\rm out}=\frac{1-T}{1+T}.
\end{eqnarray}
Comparison of the formula for $\nu_{\rm out}$ just obtained with
the formula for the noise $\nu_{\rm cl}$ given in
Eq.~(\ref{nufeed-forward}) reveals that $\nu_{\rm cl}\nu_{\rm
out}=1$ holds and therefore the third of inequalities
(\ref{uncertainty}) is saturated.

\section{Experiment}\label{sec_4}

After the theoretical part where we proved the optimality of the
scheme depicted in Fig.~\ref{scheme} we now proceed by
describing the actual experiments demonstrating minimal disturbance measurements (MDMs) on
coherent states. As mentioned above, a MDM was recently performed on
coherent states \cite{Andersen_06} in which the output signal had
the same mean value as the input.
The present work extends this previous work to a more complete
experimental study of MDMs of coherent states namely to the cases where
the mean value of the input state is not preserved.  We
systematically investigated two cases. First, we considered the
non-unity gain phase-insensitive MDM where the gain was optimized
according to the formula (\ref{gopt}). In the second case we
studied the non-unity gain MDM where the optical
gain was fixed.

The experimental setup is depicted in Fig.~\ref{scheme}. In both
experiments we used a stable continuous wave Nd:YAG
laser from Innolight oscillating at 1064 nm wavelength  which
could deliver up to 500 mW of power in one transversal mode. The
signal beam, local oscillator (LO) beam and the auxiliary beam
(B) were all obtained from this source enabling high quality
mode matching between the beams and hence allowing very efficient
quantum measurements. We generated the coherent states by placing
concatenated electro-optical phase and amplitude modulators in the
beam path.  We applied a signal to the modulators at 14.3 MHz
which created sidebands with respect to the laser carrier,
meaning that some of the photons from the carrier were transferred
to these sidebands. 
Hence we defined our coherent state to
reside at the sideband frequency of 14.3 MHz and having a 100 kHz
bandwidth. In this operating window the dark noise
of the detectors was negligible and the locking loops for the
amplitude and phase quadrature measurement at the homodyne
detector were optimized to operate stably. In addition, the
feed-forward loop was chosen to function most efficiently inside this window.
After the preparation, the coherent state impinges on a beam
splitter BS(R,T) with variable beam splitting ratio. The
variable beam splitter is realized by placing a polarizing
beam splitter behind a half wave plate in the beam path. The
reflected part of the input state is mixed with an auxiliary
beam of equal intensity on a 50:50 beam splitter. By directly measuring the output of the beam splitter and subsequently constructing the sum and difference photocurrents, the amplitude and phase quadratures of the reflected light are simultaneously measured~\cite{leuchs99}, thus information about the input is acquired. The classical measurement outcome is amplified electronically and fed to another pair of
amplitude and phase modulators which are traversed by another auxiliary beam. This beam is then coupled into the remaining part of the signal beam thus accomplihing a lossless displacement operation.
The output state $\rho$ is finally analyzed by making use of a
homodyne detector. A bright local oscillator beam interferes with the signal beam on
a beam splitter, and the conjugate quadratures, amplitude and phase, are stably measured by locking the relative phase between the two beams employing standard electronic feedback techniques. The
first and second moments of the amplitude and phase quadratures of the output state as well as the input state are thus measured using an electronic spectrum analyzer. The input state was characterized by switching off the displacement operation, measuring the resulting output state and inferring the input state by carefully characterizing all the losses including the detector and beam splitter losses. 
In all the measurements the central frequency was 14.3~MHz, the resolution bandwidth was 100~kHz and the video bandwidth was 30~Hz. 



In the first experiment the optical gain of our measurement device
depended on the transmission of the beam splitter
BS(R,T) according to Eq.~(\ref{gopt}) thereby reducing the noise
$\nu_{\rm out}$ to a minimum possible value. This was achieved by
tuning the variable beam splitter to various
transmission/reflection ratios and correspondingly adjusting the feed-forward electronic
gain to obtain the desired optimal optical gain.

\begin{figure*}[] \centering \includegraphics[width=18cm]{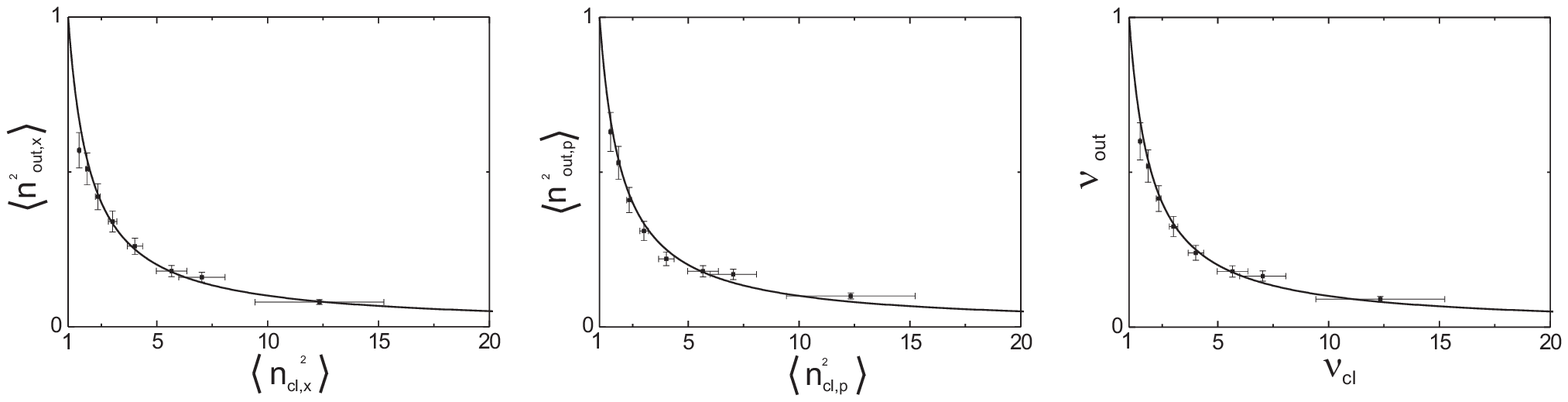}
\caption{Experimental results for MDM with optimized gain.
Variances $\langle n^2_{\rm out,x}\rangle$ (left figure) and
$\langle n^2_{\rm out,p}\rangle$ (middle figure) of the added
noises in the amplitude and phase quadratures are plotted against
the quantity $\langle n^2_{\rm cl,x}\rangle$ and $\langle n^2_{\rm cl,p}\rangle$ characterizing the noise added into
the outcomes of simultaneous measurement of amplitude and phase
quadratures. We also make use of Eq.(7) to plot $\nu_{\rm out}$ against $\nu_{\rm cl}$ (right figure). The solid line represents the theoretical relation
$\nu_{\rm out}=1/\nu_{\rm cl}$. The experimental data were
obtained by taking into account the detection efficiency of $83\%$
at the homodyne detector. The error bars in the $x$-axis steam
from the uncertainty in the measurement of the beam splitter
transmission ($2\%$ deviation). The error bars in the $y$-axis are
caused by 0.1 dB relative measurement accuracy of the Electronic
Spectrum Analyzer and 0.1 dB deviation of the homodyning
efficiency.} \label{fig3}
\end{figure*}

In order to quantify our measurement
device and to verify that it indeed measures the coherent state
optimally with minimal disturbance we needed to determine the
added noises $\langle{n^2_{\rm out,x}}\rangle$, $\langle{n^2_{\rm
out,p}}\rangle$, $\langle{n^2_{\rm cl,x}}\rangle$ and
$\langle{n^2_{\rm cl,p}}\rangle$. For this purpose we measured the
Signal to Noise Ratio of the input $\mathrm{SNR}_{\rm in}$ and the Signal
to Noise Ratio of the output $\mathrm{SNR}_{\rm out}$ for the conjugate
amplitude and phase quadratures, respectively (see e.g.~\cite{cloning}). The variances of
added noises in the output quadratures $\langle{n^2_{\rm
out,x}}\rangle$ and $\langle{n^2_{\rm out,p}}\rangle$ then read as
\begin{eqnarray}\label{SNR}
\langle{n^2_{\rm out,x}}\rangle&=&\frac{\mathrm{SNR}_{\rm in,x}}{\mathrm{SNR}_{\rm
out,x}}-1,\quad \langle{n^2_{\rm out,p}}\rangle=\frac{\mathrm{SNR}_{\rm
in,p}}{\mathrm{SNR}_{\rm out,p}}-1.\nonumber\\
\end{eqnarray}
The variances $\langle{n^2_{\rm cl,x}}\rangle$ and
$\langle{n^2_{\rm cl,p}}\rangle$ of the noises added into the
classical measurement outcomes were calculated from the measured transmittance $T$
of the variable beam splitter. By construction, the device should exhibit identical
transmittance for the amplitude and phase quadratures, and this
was explicitly confirmed by measurement. We thus have
\begin{equation}\label{nclT}
\nu_{\mathrm{cl}}=\langle{n^2_{\rm cl,x}}\rangle=\langle{n^2_{\rm cl,p}}\rangle
=\frac{1+T}{1-T}.
\end{equation}
Simultaneously changing the transmittance T
and the electronic gain G enabled us to adjust at will the degree of
disturbance of the measured quantum state. The experimental
results are summarized in Fig.~\ref{fig3}. We get excellent agreement between theory and
experiment and we conclude that the measurement apparatus operates at the fundamental limits imposed by quantum theory.

In our second experiment we demonstrated the MDM for a fixed
optical gain. After analyzing our input state as in the previous
experiment we connected the feed-forward loop and adjusted the
appropriate electronic gain which guaranteed the desired fixed
optical gain. This means that for a particular beam splitter
transmission the feed-forward electronic gain will increase as the
desired optical gain increases. If, in addition, the desired
optical gain is less than the beam splitter transmission then a
deamplification of the optical signal is required which was
achieved by adding a $\pi$ phase shift in the electronic
feed-forward loop and by making use of destructive interference
which resulted in optical deamplification.  This particular operation is actually suboptimal meaning that the measurement outcomes lie outside the optimality window which follows directly from the fact that in this case where $g<\sqrt{T}$, calculation of $\nu_{\rm cl}$ reveals that $\nu_{\rm cl}>(1+g^2)/|1-g^2|$. Using a similar
procedure as before we measured the first and second moments for
the amplitude and phase quadrature of the output signal. Having
measured the input and output states we then calculate all the
necessary added noises by means of Eqs.~(\ref{SNR}) and
(\ref{nclT}).
\begin{figure*}[]


\centering\includegraphics[width=18cm]{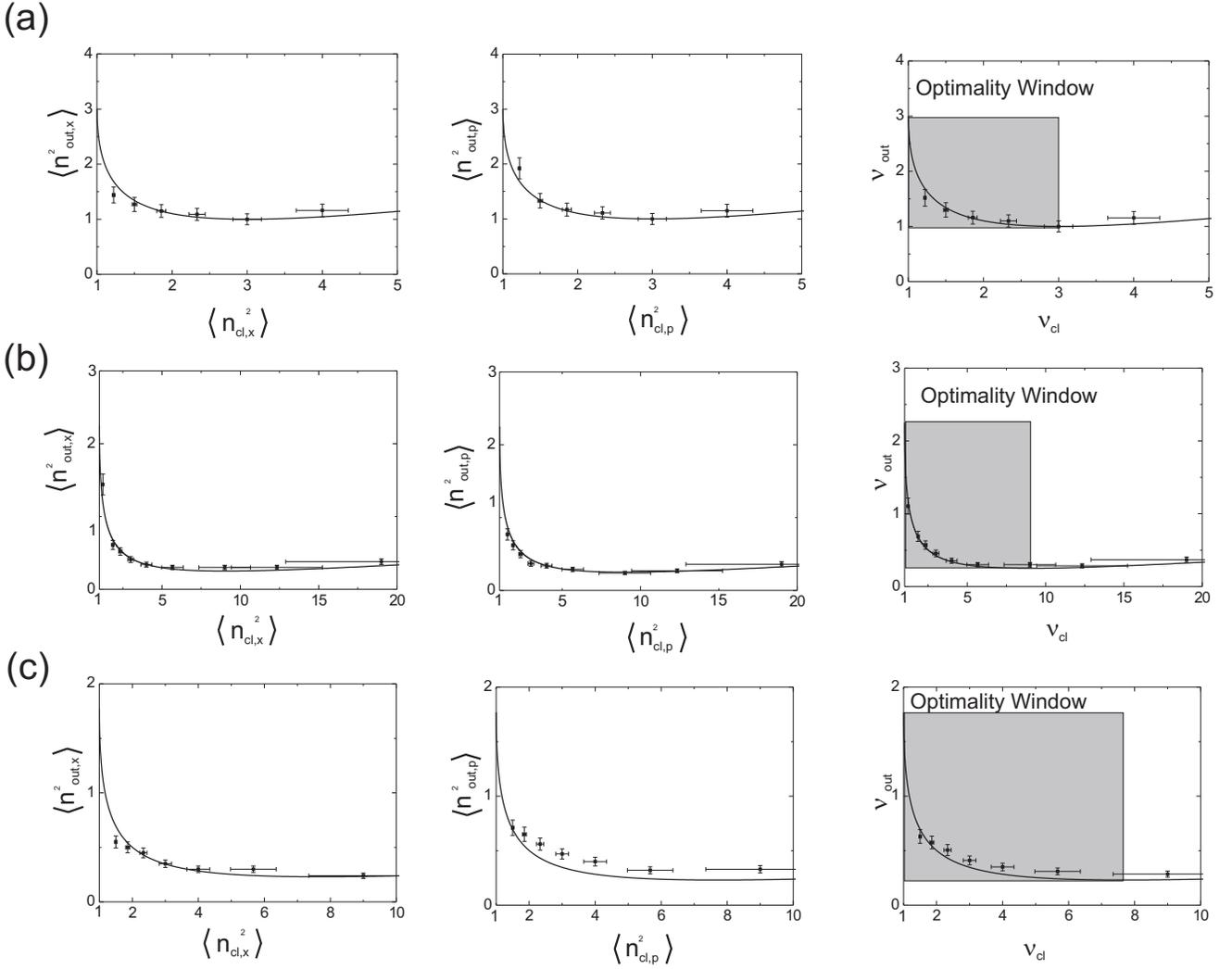} \caption{
Experimental results for MDM with fixed gains $g^2=0.5$ (a),
$g^2=0.8$ (b) and $g^2=1.3$ (c). Variances $\langle n^2_{\rm out,x}\rangle$ (left figure) and $\langle n^2_{\rm out,p}\rangle$ (middle figure) of the added
noises in the amplitude and phase quadratures are plotted against
the $\langle n^2_{\rm cl,x}\rangle$ and $\langle n^2_{\rm cl,p}\rangle$ characterizing the noise added into the outcomes of simultaneous measurement of amplitude and phase
quadratures. We also make use of Eq.(7) to plot $\nu_{\rm out}$ against $\nu_{\rm cl}$ (right figure). The optimality
windows (gray shaded regions) are determined by
Eqs.~(\ref{inequality1}) and (\ref{inequality2}). The solid line
represents the theoretical trade-off (\ref{trade-off}). The
experimental data were obtained by taking into account the
detection efficiency of $83\%$ at the homodyne detector. The error
bars in the $x$-axis stem from the uncertainty in the measurement
of the beam splitter transmission ($2\%$ deviation). The error
bars in the $y$-axis are caused by 0.1 dB relative measurement
accuracy of the Electronic Spectrum Analyzer and 0.1 dB deviation
of the homodyning efficiency.} \label{fig4}
\end{figure*}

We performed this kind of measurement for three fixed optical gains
of  $g^2=0.5,0.8,1.3$ and the results are summarized in
Fig.~\ref{fig4}. Here we have to stress again that optimal MDMs with a fixed
optical gain lie only within a specific region of the
$(\nu_{\rm cl},\nu_{\rm out})$-plane (gray shaded region in the
figures). The boundaries of these regions depend solely on the optical gain and
can be easily determined from Eqs.~(\ref{inequality1}) and
(\ref{inequality2}). For the optical gains considered here these
optimality windows explicitly read as follows: 
\begin{eqnarray*}
&1\leq \nu_{\rm cl}\leq 3,&\quad
1\leq \nu_{\rm out}\leq 3\quad \mbox{for $g^{2}=0.5$},\\
&1\leq \nu_{\rm cl}\leq 9,&\quad \frac{1}{4}\leq \nu_{\rm out}\leq
\frac{9}{4}\quad
\mbox{for $g^{2}=0.8$},\\
&1\leq \nu_{\rm cl}\leq \frac{23}{3},&\quad \frac{3}{13}\leq
\nu_{\rm out}\leq \frac{23}{13}\quad \mbox{for $g^{2}=1.3$}.
\end{eqnarray*}
The left hand sides of these inequalities are dictated by the
commutation relations (\ref{commutators}) and cannot be overcome
by any operation, i.e. the theoretical trade-off in the figures
never lies below or to the left of the optimality window.
However, the trade-off (\ref{trade-off}) lies to the right of the
optimality window for sufficiently large noise $\nu_{\rm cl}$. In
this case the operation saturating the trade-off is suboptimal and
a better performance is obviously obtained by the
operation corresponding to the minimum of the trade-off which also corresponds to the bottom right corner of the optimality window.  Note that the only exception occurs in the
unity gain regime ($g=1$) demonstrated in \cite{Andersen_06} where
the optimality window is not bounded from the right, i.e.
\begin{equation}\label{inequality3}
1\leq \nu_{\rm cl}\leq \infty,\quad 0\leq \nu_{\rm out}\leq 2.
\end{equation}
Thus, since $\nu_{\rm cl}-\sqrt{\nu_{\rm cl}^{2}-1}\leq 1$ for any
$\nu_{\rm cl}\geq 1$ the optimal output noise $\nu_{\rm out}$ for
$g=1$ always satisfies the second inequality
in (\ref{inequality3}) and therefore in the unity gain regime the
trade-off (\ref{trade-off}) never leaves the optimality window.

As can be seen in Fig.~\ref{fig4}, the obtained experimental trade-offs
are in very good agreement with the theory which shows that our measuring apparatus
indeed realizes optimal non-unity gain Gaussian partial estimation of coherent states.
In particular, the noises added to the phase and
amplitude quadratures are practically the same, which confirms that
the measurement procedure introduces isotropic phase-independent noise into the estimated
state as well as the post-measurement state.

\section{Discussion and conclusions}\label{sec_6}
Our experimental minimal disturbance measurement with fixed non-unity gain
finds a direct application in the context of optimal Gaussian individual
attacks on coherent state quantum key distribution (QKD) with heterodyne
detection and direct reconciliation \cite{Weedbrook_04}. The optimal trade-off
demonstrated by us determines the minimum added noise in the outcomes
of simultaneous measurement of complementary quadratures a
potential eavesdropper can reach for a Gaussian quantum channel
with a fixed gain and a fixed phase-insensitive added noise.
In the QKD terminology it means that the minimal disturbance measurement
provides an eavesdropper with maximum possible information that can be gained from an individual
Gaussian attack in the heterodyne-based coherent state QKD protocol with direct
reconciliation.
Recently, the similar problem has been studied theoretically directly
in the context of QKD \cite{Lodewyck_07} and another form of the above
mentioned
trade-off was found (Eq.~(11) of Ref.~\cite{Lodewyck_07}). The proofs of
optimality presented here,
however, follow completely different strategies in comparison
with those presented in \cite{Lodewyck_07} and more importantly
we saturate the optimal trade-off between added noises experimentally.

In this article we have extended the concept of the
phase-insensitive MDM for coherent states to the non-unity gain
regime. We have given a complete theoretical as well as
experimental study of this MDM for two different scenarios. First,
we have found the non-unity gain MDM assuming fixed optical gain.
In the second scenario we considered MDM with optimized gain. We
have shown that both MDMs can be realized by a scheme
consisting of only linear optical elements and a feed-forward and
we implemented the scheme experimentally. We have experimentally
reached theoretical limits in both scenarios. Our results give
answer to a fundamental question of how much noise will be in the
measurement outcomes from the non-destructive
measurement of a coherent state provided that it is represented by
a single-mode Gaussian channel with a given optical gain and
isotropic added noise. Our analysis could be also extended to
phase-sensitive measurements such as the quantum non-demolition measurement
of a single quadrature of light.

\acknowledgments

The research has been supported by the research projects
``Measurement and Information in Optics,'' (MSM 6198959213) and
Center of Modern Optics (LC06007) of the Czech Ministry of
Education. Support by the COVAQIAL (FP6-511004) and SECOQC
(IST-2002-506813) projects of the sixth framework program of EU is
also acknowledged. R. F. acknowledges support from Alexander von
Humboldt Fellowship.


\end{document}